\documentclass{article}
\usepackage{spconf,amsmath,graphicx,hyperref}
\usepackage{booktabs}
\usepackage{multirow}
\usepackage{makecell}
\usepackage{color,soul}
\usepackage{xcolor}
\usepackage{flushend}

\title{Multi-Channel Differential ASR for Robust Wearer Speech Recognition on Smart Glasses}
\name{\begin{tabular}{c}Yufeng Yang$^{1, 2*}$\thanks{*Work done during internship at Meta}, Yiteng Huang$^{2}$, Yong Xu$^{2}$, Li Wan$^{2}$, Suwon Shon$^{2}$, Yang Liu$^{2}$, Yifeng Fan$^{2}$, \\
Zhaojun Yang$^{2}$, Olivier Siohan$^{2}$, Yue Liu$^{2}$, Ming Sun$^{2}$, Florian Metze$^{2}$\end{tabular}}
\address{$^{1}$The Ohio State University, USA \quad $^{2}$Meta, USA\\
\texttt{yang.5662@osu.edu, yah@meta.com, yongxu@meta.com}}

\begin{document}
\ninept

\maketitle

\begin{abstract}
With the growing adoption of wearable devices such as smart glasses for AI assistants, wearer speech recognition (WSR) is becoming increasingly critical to next-generation human-computer interfaces. However, in real environments, interference from side-talk speech remains a significant challenge to WSR and may cause accumulated errors for downstream tasks such as natural language processing. In this work, we introduce a novel multi-channel differential automatic speech recognition (ASR) method for robust WSR on smart glasses. The proposed system takes differential inputs from different frontends that complement each other to improve the robustness of WSR, including a beamformer, microphone selection, and a lightweight side-talk detection model. Evaluations on both simulated and real datasets demonstrate that the proposed system outperforms the traditional approach, achieving up to an 18.0\% relative reduction in word error rate.

\end{abstract}


\section{Introduction}
Automatic speech recognition (ASR) systems have seen remarkable progress in recent years, driven by advances in deep learning and end-to-end (E2E) neural network architectures~\cite{hinton2012deep, lecun2015deep, prabhavalkar2023end}. Despite these advancements, wearer speech recognition (WSR) on wearable devices such as smart glasses remains underexplored. As a critical part of an emerging next-generation human-computer interface~\cite{lee2018interaction, engel2023project}, WSR on smart glasses requires microphones to remain active for long and continuous interactions. Unlike traditional close-talk systems, microphones on smart glasses operate in open-field conditions, making WSR particularly vulnerable to bystander side-talk in real environments. Improving WSR robustness is essential to ensure a reliable user experience on smart glasses and other wearable devices.

Smart glasses are equipped with multiple microphones, enabling the integration of a beamformer in an ASR system for smart glasses. Non-linearly constrained minimum variance (NLCMV) beamforming~\cite{lin2024agadir} was proposed for smart glasses, which incorporates white noise gain and null direction control, to project multi-channel microphone inputs into predefined directions. This technique was adopted in several subsequent studies~\cite{zmolikova2024chime, feng2025directional, yang2025m, jiang2025multi, lan2026exploring, lin2025directional, xie2025thinking} for conversational ASR on smart glasses. However, when only recognizing the wearer in the presence of a bystander, the traditional approach cannot fully suppress the degradation caused by the side-talk in WSR.

Traditional frontends such as speech enhancement~\cite{loizou2007speech} and talker-independent speaker separation~\cite{wang2018supervised} can effectively improve the speech intelligibility and quality. However, even with causal models, the added latency by integrating such models makes it impractical for real applications. Other frontend methods, such as speaker diarization~\cite{park2022review} and target speaker extraction~\cite{zmolikova2023neural}, have the same problem. Moreover, the application of WSR on smart glasses constrains frontend design, since modeling speaker information may raise privacy concerns. Consequently, a new ASR system design is required to improve WSR performance on smart glasses.

In this work, we propose multi-channel differential ASR, a novel method that takes inputs from different frontends that complement each other, for robust WSR on smart glasses. Unlike the traditional approach, we incorporate frontends alongside a beamformer to provide complementary spatial cues for the ASR model. We utilize microphone selection to choose the channel with the highest signal-to-noise ratio (SNR), thereby minimizing additional latency. We also integrate a lightweight streaming side-talk detection (STD) model to distinguish between the wearer and bystander without modeling speaker identity. For the beamformer, we utilize an adjusted minimum variance distortionless response (MVDR) beamformer. The ASR backbone is based on a low-latency streaming recurrent neural network transducer (RNN-T) network~\cite{graves2012rnnt}. Evaluations on both simulated and real recorded datasets demonstrate that combining microphone selection, beamforming, and STD outperforms the traditional approach that relies solely on beamforming as the frontend, with up to 18.0\% relative reduction in word error rate (WER).

We make several contributions to the field. We propose a novel multi-channel differential ASR for robust WSR. With a careful design, the outputs from different frontends with different frame rates can be combined. We also create a real recorded dataset to analyze the WSR performance for different bystander angles, distances, and heights. The proposed differential ASR system has the potential to be extended to other ASR applications as well.


\section{Proposed Method}\label{sec:system}

\begin{figure*}[!t]
    \centering
    \includegraphics[width=0.99\linewidth]{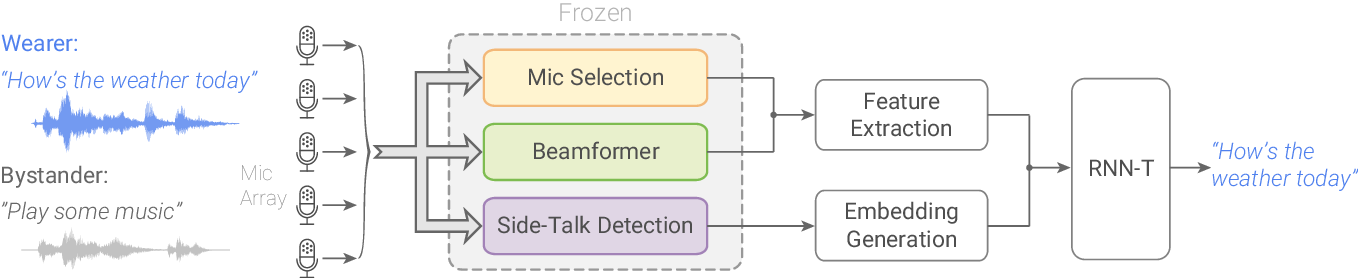}
    \caption{Diagram of the proposed multi-channel differential ASR system for robust WSR on smart glasses.}
    \label{fig:diagram}
\end{figure*}

\subsection{Microphone Array Beamforming}
Microphone array beamforming is a spatial filtering technique in speech processing that leverages a microphone array to enhance the speech signal from desired directions while suppressing noise and interference from other spatial locations~\cite{van1988beamforming, benesty2008microphone}. The technique works by applying filter weights to signals from multiple microphones, effectively steering the array's sensitivity toward the target speech source while suppressing unwanted acoustic sources. Traditional beamforming approaches include delay-and-sum beamforming, MVDR beamforming, and others~\cite{van1988beamforming, benesty2008microphone, huang2025advances}. For smart glasses, NLCMV beamformer~\cite{lin2024agadir} has shown good performance on ASR. In this work, we adopt an internal MVDR-based beamformer that directs the beam to the wearer's mouth. This beamformer is more suitable for WSR than NLCMV because it focuses solely on the wearer.

\subsection{RNN-T Based Streaming ASR}
For wearable devices such as smart glasses, a low-latency streaming ASR model is essential to ensure continuous transcription throughout extended usage sessions. The RNN-T model~\cite{graves2012rnnt} is particularly well suited for this task, as it enables fully streaming, E2E processing with three integrated components: an acoustic encoder that extracts features from audio input, a prediction network that functions like an internal language model, and a joint network that merges their outputs to generate predictions. In this work, we build our RNN-T on the Emformer~\cite{shi2021emformer} architecture, which features an efficient memory Transformer designed for low-latency streaming ASR. 


\subsection{Side-Talk Detection}
STD is a novel task for WSR on smart glasses. It distinguishes between wearer and bystander speech without modeling speaker identity, thereby protecting privacy. The task was introduced in~\cite{liu2025mmw} to build a Whisper~\cite{radford2023robust} model robust to side-talk speech. In this work, we design a streaming STD model that operates at the audio sample level and outputs logit scores indicating voice activity from the wearer, bystander, or non-speech segments. The STD model is based on a temporal convolutional network (TCN)~\cite{bai2018tcn}, and is lightweight, with approximately 2M parameters.

\subsection{Multi-Channel Differential ASR}
We propose multi-channel differential ASR, a novel ASR system designed to improve the robustness of WSR to side-talk. Unlike the traditional approach, where the ASR model relies solely on a beamformer as the frontend to process multi-channel microphone inputs, differential ASR leverages different frontend modules that provide complementary or contrastive information to each other. The diagram of the proposed differential ASR is shown in Fig.~\ref{fig:diagram}. 

In the proposed differential ASR system, we adopt a microphone selection module, a beamformer, and an STD model. The microphone selection module chooses the channel with the highest SNR from the input microphone array. The selection is fixed and based on the microphones' physical location relative to the wearer's mouth. Given a pair of smart glasses, the closest microphone to the wearer's mouth will be selected. Thus, we denote the output of the microphone selection as \textbf{ch-0}. For the beamformer, the internal modified MVDR beamformer is utilized, which takes all microphone signals as input and generates a single-channel beamformed audio as output, which we denote as \textbf{ch-x}. Based on the STD model logits, we generate an embedding (denoted as \textbf{embed}), which is concatenated with the log-Mel feature of ch-0 and ch-x as input to the RNN-T model. All frontends are frozen, so the number of additional trainable parameters remains under 1M compared to the traditional system.

\section{Experimental Setup}\label{sec:exp}

\begin{figure}[b!]
    \centering
    \includegraphics[width=0.8\linewidth]{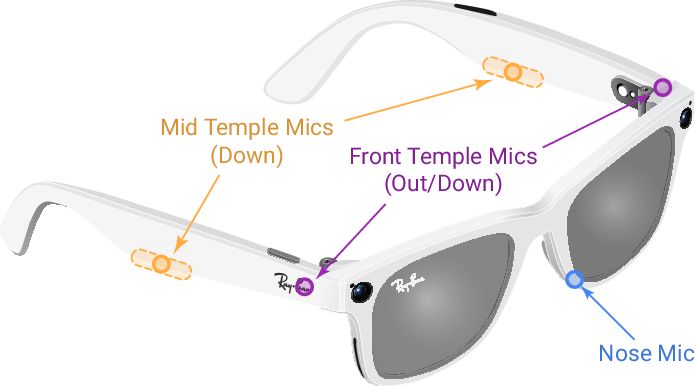}
    \caption{Microphone location on a pair of smart glasses.}
    \label{fig:sg}
\end{figure}

\subsection{Datasets}
We designed a setup to evaluate the performance of robust WSR on smart glasses. The experiments are based on the LibriSpeech~\cite{panayotov2015librispeech} dataset. We first simulate multi-channel LibriSpeech using room impulse responses (RIRs) measured on a pair of Ray-Ban Meta smart glasses. The microphone location of the smart glasses is shown in Fig~\ref{fig:sg}. There are 5 microphones in total, with one near the nose and two pairs of microphones on the front and mid temples. We prepare two categories of datasets: wearer-only (clean) and wearer with side-talk (noisy). The clean data does not contain noise or side-talk speech. For the simulated noisy data, side-talk speech from a bystander is added as noise, with wearer-to-bystander SNRs ranging from 10 to 25 dB. We prepare a clean training set and a noisy training set with 500554 utterances each. The wearer and bystander speech are sourced from all LibriSpeech training sets. In each wearer and bystander mixture, bystander speech is randomly sampled other than the wearer speech, and added to the wearer speech with a random overlap ratio from 0\% to 100\%. Validation data is clean only, sourced from all LibriSpeech validation sets with 6747 utterances. For simulated evaluation, we test on both clean and noisy test-clean and test-other sets, with 3558 and 3502 utterances, respectively. In each noisy test set, two overlap ratios of 0\% and 50\% are utilized.

\begin{figure}
    \centering
    \includegraphics[width=\linewidth]{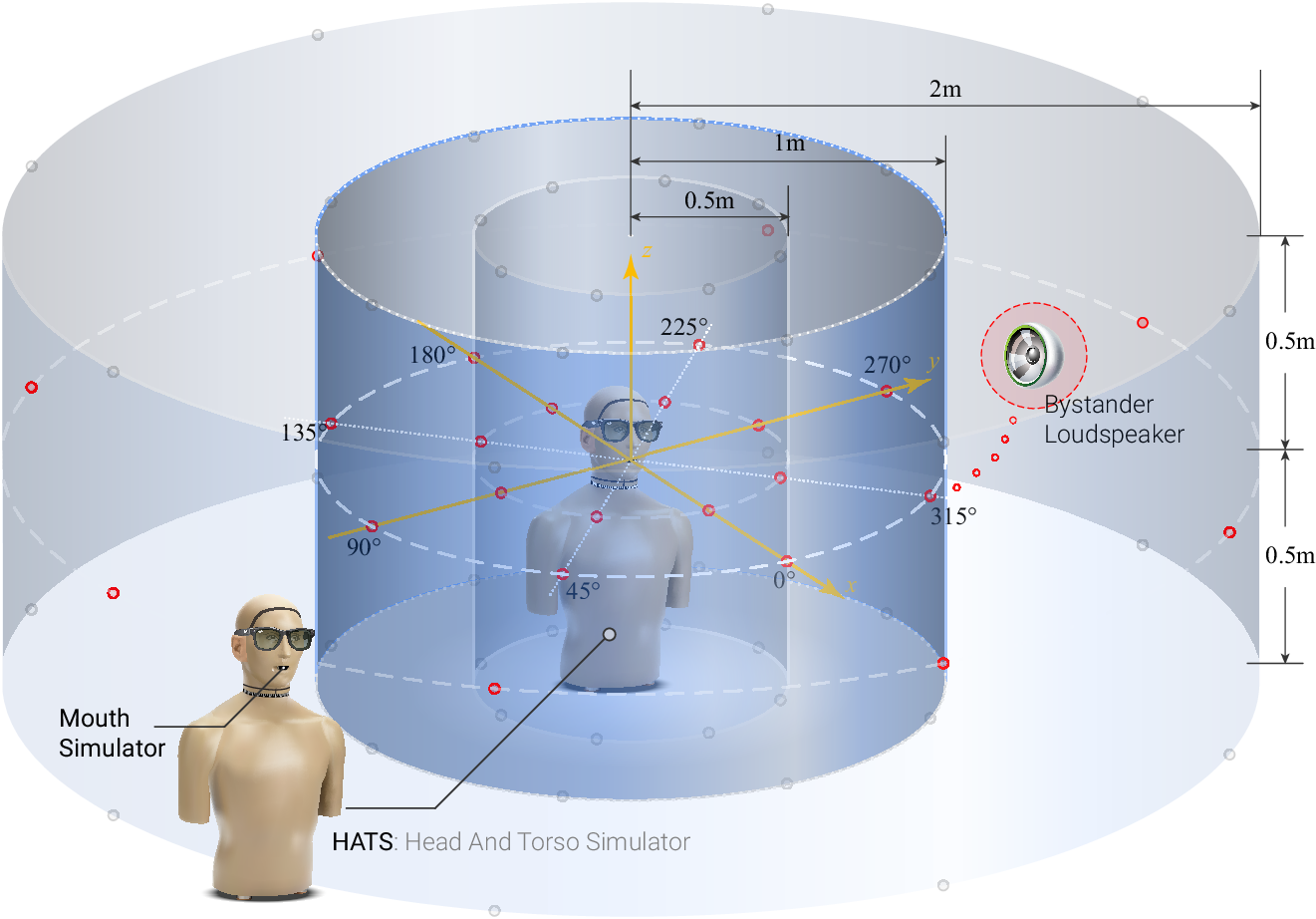}
    \caption{Recording setup for data collection with HATS.}
    \label{fig:hats_collection}
\end{figure}

To evaluate the WSR performance in real environments, we set up a head and torso simulator (HATS) and loudspeakers to collect a real dataset for evaluation. A pair of Ray-Ban Meta smart glasses is mounted on the HATS to capture multi-channel data. The wearer's speech is played back via the HATS mouth simulator, and the bystander speech is played from loudspeakers placed around the HATS, as shown in Fig.~\ref{fig:hats_collection}. The bystander loudspeaker covers 72 distinct locations, including 8 different angles of \{0, 45, 90, 135, 180, 225, 270, 315\} degrees to the HATS, 3 different relative heights of \{-0.5, 0.0, 0.5\} m to the wearer, and 3 different distances of \{0.5, 1.0, 2.0\} m to the wearer. The wearer and bystander speech are sourced from test-clean and test-other sets, respectively. The wearer and side-talk speech are recorded separately and mixed during post-processing. The mixing process follows the same procedure as for the simulated test data. On real data, we distinguish the order of the first speaker. Either the wearer speaks first (wearer-bystander) or the bystander speaks first (bystander-wearer). For each speaker order and overlap ratio, the resulting evaluation set has 188640 utterances for all 72 bystander locations.

\subsection{Implementation Details}

Based on the physical locations of the microphones on the smart glasses, the nose microphone is closest to the wearer's mouth, and is selected as ch-0 in the microphone selection module. The beamformer takes in all 5-ch inputs and generates a 1-ch beamformed audio. The STD model is trained on real non-user data and produces logits for the wearer at the audio sample level. As shown in Fig.~\ref{fig:diagram}, only uncolored modules are trainable and all frontends are frozen. 

The feature extraction module extracts an 80-dimensional log-Mel feature when only ch-x is fed as input. When ch-x is coupled with ch-0, the feature extraction module concatenates their log-Mel features and uses two streaming 2D convolutional layers (Conv2D) with a kernel size of [2, 5], a stride of [1, 2], and a gated linear unit (GLU) activation function. The log-Mel feature of each channel is halved in dimension, then concatenated as the output of the feature extraction module. The embedding generation module contains two Conv2D layers with a kernel size of [20, 1] and a stride of [10, 1] and [16, 1] for each layer, respectively. After this processing, the frame rate matches the output of the feature extraction module. The intermediate number of channels is 3, and the final embedding dimension is 5. Afterwards, the extracted ch-x and ch-0 features and embed are concatenated and fed to the RNN-T model. All Conv2D layers in the feature extraction and embedding generation module are followed by a 2D batch normalization layer.

The RNN-T model first reduces the input frame size by 6. Then, 20 layers of Emformer are configured with an input dimension of 320, number of heads of 4, feedforward dimension of 2048, context of 10 past frames, segment size of 2, activation function of Gaussian error linear unit (GELU), and a convolutional kernel of [7, 0] with Swish activation. The encoder output dimension is 768. The output units are 4096 sentence pieces~\cite{kudo2018sentencepiece} with byte pair encoding (BPE)~\cite{vesely2013bpe} as the segmentation algorithm. In the predictor, the tokens are first represented by 256-dimensional embeddings and processed by two long short-term memory (LSTM) layers with 256 hidden nodes, followed by a linear projection to a 768-dimensional feature. The joint network projects the input to the number of sentence pieces. 

We built five systems for comparison. Following the traditional approach, we only take the beamformer as the frontend and use ch-x for ASR to serve as our baselines. Two baselines are trained, one on clean data and the other on noisy data, denoted by clean-trained ch-x and noisy-trained ch-x, respectively. For the proposed differential ASR system, we try different combinations of ch-x, ch-0, and embed, and denote them as ch-x + embed, ch-x + ch-0, and ch-x + ch-0 + embed, respectively representing the system leveraging beamformer and STD model, beamformer and microphone selection, and all three frontends. All differential ASR systems are trained on the noisy data since side-talk resistance is our main focus. For all five systems, the trainable parameters are $\sim$70M with a 120 ms ASR latency. All models are trained on 32 NVIDIA H100 GPUs with a batch size of 3600. The Adam optimizer was used with betas of (0.9, 0.98), epsilon of 1$e^{-8}$, and weight decay of 1$e^{-6}$. A tri-stage learning rate schedule was used with a peak learning rate of 0.0005, warmed up for 20k steps. The models are trained with the RNN-T loss, and the final checkpoint is selected based on the validation WER.

\section{Results and Discussion}\label{sec:result}

\begin{table*}[ht!]
    \centering
    \caption{ASR (\%WER) results on the simulated multi-channel LibriSpeech test set.}
    \label{tab:simu}
    \centering
    \begin{tabular}[width=\linewidth]{lcccccccc}
        \toprule
         \multirow{2}{*}{\textbf{System}} & \multicolumn{3}{c}{\textbf{Wearer-only}} & \multicolumn{5}{c}{\textbf{Wearer with Side-talk}} \\
         \cmidrule(l{0.5em}r{0.5em}){2-4}
         \cmidrule(l{0.5em}r{0.5em}){5-9}
          & test-clean & test-other & Avg & \multicolumn{2}{c}{test-clean} & \multicolumn{2}{c}{test-other} & Avg \\
          \midrule
          OVERLAP RATIO & - &-&-& 0\% & 50\% & 0\% & 50\% & - \\
         \midrule
         
          Clean-trained ch-x & \textbf{5.70} & \textbf{14.75} & \textbf{10.23} & 88.62 & 46.07 & 89.75 & 51.23 & 68.92 \\
          Noisy-trained ch-x & 6.46 & 16.48 & 11.68 & 6.37 & 6.63 & 16.72 & 17.42 & 11.79   \\ 
          \midrule
          Noisy-trained ch-x + embed & 6.06 & 15.99 & 11.14 & \textbf{5.98} & \textbf{6.20} & \textbf{16.01} & \textbf{16.57} & \textbf{11.19} \\
          Noisy-trained ch-x + ch-0& 6.21 & 16.30 & 11.39 & 6.07 & 6.46 & 16.34 & 16.96 & 11.46 \\
          Noisy-trained ch-x + ch-0 + embed & 6.07 & 16.08 & 11.21 & 6.03 & 6.21 & 16.11 & 16.78 & 11.28 \\
         
        \bottomrule
    \end{tabular}
\end{table*}

\begin{table*}[htbp!]
    \centering
    \caption{ASR (\%WER) results on the real recorded multi-channel LibriSpeech test set.}
    \label{tab:real}
    \centering
    \begin{tabular}[width=\linewidth]{lcccccc}
        \toprule
         \multirow{2}{*}{\textbf{System}} & \multirow{2}{*}{\textbf{Wearer-only}} & \multicolumn{5}{c}{\textbf{Wearer with Side-talk}}\\
         \cmidrule(l{0.5em}r{0.5em}){3-7}
         & & \multicolumn{2}{c}{wearer-bystander} & \multicolumn{2}{c}{bystander-wearer} & Avg \\
          \midrule
         OVERLAP RATIO & - & 0\% & 50\% & 0\% & 50\% & - \\
         \midrule
         
          Clean-trained ch-x & 6.30 & 29.20 & 15.28 & 40.96 & 23.81 & 27.31 \\
         Noisy-trained ch-x & 7.20 & 7.19 & 7.41 & 7.22 & 7.63 & 7.36 \\
         \midrule
         Noisy-trained ch-x + embed & 6.82 & 6.79 & 7.02 & 6.85 & 7.06 & 6.93 \\
          Noisy-trained ch-x + ch-0 & 6.51 & 6.50 & 6.57 & 6.38 & 6.50 & 6.49 \\
          Noisy-trained ch-x + ch-0 + embed & \textbf{6.29} & \textbf{6.28} & \textbf{6.37} & \textbf{6.30} & \textbf{6.26} & \textbf{6.30} \\
        \bottomrule
    \end{tabular}
\end{table*}

\subsection{Results on Simulated Data}

Evaluation results on simulated data are presented in Table~\ref{tab:simu}. We compare five systems on different test data. Clean-trained ch-x and noisy-trained ch-x serve as our baselines, and different combinations of ch-0 and embed with ch-x are investigated. The clean-trained ch-x performs the best on all clean test sets with 10.23\% WER on average, because of the matched training condition. All proposed systems outperform the noisy-trained ch-x. When evaluating on the noisy data, the performance of the clean-trained ch-x degrades significantly due to the side-talk speech. The noisy-trained ch-x outperforms the clean-trained ch-x with an 82.8\% relative WER reduction (WERR) on average, which shows the importance of training data augmentation. Trained on the same data, the proposed systems outperform noisy-trained ch-x, and the best results come from ch-x + embed, with 5.1\% relative WERR. The results highlight the effectiveness of both data augmentation and the differential ASR system compared to the traditional approach.

\subsection{Results on Real Recorded Data}
Table~\ref{tab:real} shows the WER comparison on the real data. For noisy test data, we present the average WER across all 72 bystander heights, angles, and distances. On clean data, clean-trained ch-x achieves 6.30\% WER, which is expected to perform the best. However, the proposed system ch-x + ch-0 + embed achieves a comparable 6.29\% WER, demonstrating that even trained on noisy data, the differential ASR system can perform even better than the clean-trained model on clean data, showing the effectiveness of the proposed system. All proposed systems outperform noisy-trained ch-x. On noisy test data, clean-trained ch-x has a lower WER compared with that on simulated noisy data. The noisy-trained ch-x outperforms the clean-trained ch-x with a 73.1\% relative WERR on average. All proposed systems outperform noisy-trained ch-x, and we notice ch-x + ch-0 + embed outperforms ch-x + ch-0, which outperforms ch-x + embed. This result suggests that for the RNN-T model, embed contains complementary information to ch-x, and so does ch-0. Moreover, ch-x + ch-0 outperforms ch-x + embed by 6.3\% relatively on side-talk data, indicating ch-0 contains more differential information than embed to ch-x. Lastly, by combining all three inputs, the system's performance is further elevated, showing that ch-0 and embed have contrastive information to each other. On average, our best system outperforms the strong noisy-trained ch-x baseline with an average 14.4\% relative WERR, reaching up to 18.0\%. The improvement is larger than that on simulated data because real data matches the training condition of the STD model. It is worth noting that, in Table~\ref{tab:simu} and~\ref{tab:real}, on noisy data with 0\% overlap, some WERs are lower than those on clean data, which is due to their matched train--test conditions.

\begin{figure}[htbp!]
    \centering
    \includegraphics[width=0.82\linewidth]{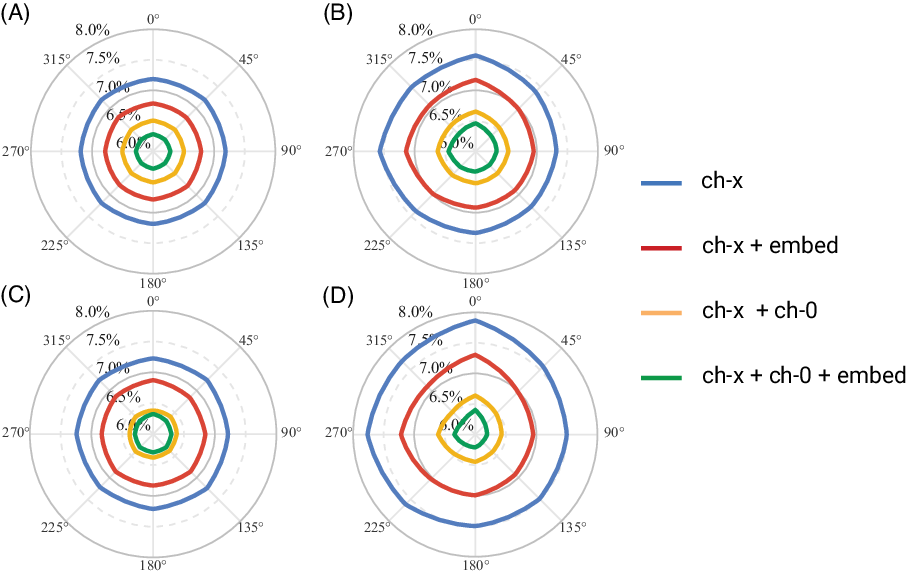}
    \caption{WER comparison on real data for different bystander angles on: (A) wearer-bystander 0\% overlap; (B) wearer-bystander 50\% overlap; (C) bystander-wearer 0\% overlap; (D) bystander-wearer 50\% overlap. All systems are trained on noisy data.}
    \label{fig:hats_lab_eval}
\end{figure}

To better understand the model's performance from different angles, we visualize the average WER across all bystander distances and heights on different angles in the noisy test data in Fig.~\ref{fig:hats_lab_eval}. In each plot, we compare four ASR systems in 8 different angles with WER ranging in [6.0\%, 8.0\%], denoted by distinct colors. With 0\% overlap, the WERs from different angles are relatively stable. However, with 50\% overlap, WERs of angle 270$^{\circ}$, 315$^{\circ}$, and 0$^{\circ}$ of wearer-bystander are higher than other angles, and for bystander-wearer, 225$^{\circ}$ is also challenging for all systems. This finding suggests a new direction for beamformer design to reduce the performance gap between different bystander angles.

\section{Concluding Remarks}\label{sec:conclusion}
In this work, we focus on the challenging side-talk problem in ASR for robust WSR on smart glasses, and propose differential ASR, a novel design that leverages different frontends that complement each other. We record a real dataset with HATS and loudspeakers to evaluate the proposed systems. Our best system utilizes beamforming, microphone selection, and an STD model. Through this integration, the proposed system outperforms a strong baseline with up to 18.0\% relative WERR, demonstrating the effectiveness of the proposed system. The concept of differential ASR can be extended to other ASR tasks with different frontends. Future work includes improving the STD model against noise and additional bystanders, deploying the proposed system on devices, and developing a frontend to better handle challenging angles for robust WSR.




\clearpage
\bibliographystyle{IEEEbib}
\bibliography{refs}

\end{document}